\documentclass[a4paper,aps,prd,10pt,preprintnumbers,nopacs,twocolumn,superscriptaddress,nofootinbib,amsmath,amssymb,preprint]{revtex4}
\usepackage{graphicx,cmap,ulem}
\usepackage[utf8]{inputenc}
\usepackage[T1]{fontenc}

\def\re#1{\mathrm{Re}(#1)}
\def\im#1{\mathrm{Im}(#1)}

\begin{document}

\title{Long-lived modes and grey-body factors of massive fields in quantum-corrected (Hayward) black holes}
\author{Alexey Dubinsky}
\email{dubinsky@ukr.net}
\affiliation{University of Seville, Seville, Spain}

\begin{abstract}
We study the dynamics of a massive scalar field in the background of the Hayward black hole, which can be interpreted both as a regular spacetime and as an effective geometry arising from Asymptotically Safe gravity.  
The quasinormal spectrum and grey-body factors are computed using the WKB method with Padé improvements and confirmed through time-domain integration followed by Prony analysis.  
We find that the mass of the field significantly suppresses the damping rate of quasinormal oscillations, giving rise to long-lived modes that continuously approach arbitrarily long-lived states (quasi-resonances) at certain critical field masses.  
In the time domain, the standard exponentially decaying ringdown is replaced by oscillatory tails with a power-law envelope.  
The corresponding grey-body factors reveal a pronounced shift of the transmission peak toward higher frequencies and a suppression of the low-frequency part of the spectrum.  
Finally, we show that the correspondence between quasinormal modes and grey-body factors remains valid for massive fields, being highly accurate for large multipole numbers and gradually losing precision as either the field mass increases or the multipole number decreases.  
\end{abstract}

\maketitle

\section{Introduction}

Quasinormal modes (QNMs) and grey-body factors (GBFs) constitute two fundamental characteristics of black-hole spacetimes.  
The QNMs describe the damped oscillations that dominate the gravitational or field perturbations at intermediate times and encode direct information about the background geometry through the complex spectrum of characteristic frequencies \cite{Kokkotas:1999bd,Berti:2009kk,Konoplya:2011qq,Bolokhov:2025uxz}.  
Their real parts correspond to oscillation frequencies, while the imaginary parts determine the damping rates.  
These modes are expected to be measurable in the ringdown phase of black-hole mergers and, therefore, provide an important probe of strong-field gravity.  
Complementarily, the GBFs determine the transmission probability of waves through the curvature-induced potential barrier and thus quantify deviations from pure blackbody radiation in Hawking evaporation \cite{Page:1976df}.  
Together, QNMs and GBFs form a consistent description of black-hole response to perturbations and play a central role in testing gravity theories and the possible quantum nature of spacetime.

\vspace{1em}
Among the various regular black-hole geometries proposed in the literature, the Hayward spacetime \cite{Hayward:2005gi} occupies a particularly important position.  
Originally introduced as a phenomenological model of a nonsingular evaporating black hole, it has later been shown that the Hayward metric naturally arises within the framework of Asymptotically Safe (AS) gravity \cite{Bonanno:2000ep,Bonanno:2016dyv,Platania:2019kyx} as an effective description incorporating quantum corrections to the Schwarzschild geometry \cite{Held:2019xde}.  
In this context, the Hayward black hole can be regarded as a self-consistent semiclassical configuration, in which the central curvature singularity is replaced by a de Sitter core sustained by quantum-gravity effects.  
The resulting metric provides a valuable theoretical laboratory for studying the impact of such quantum corrections on classical observables, including quasinormal spectra, grey-body factors, and Hawking radiation.

\vspace{1em}
An equally important aspect of black-hole perturbation theory concerns the mass of the perturbing field.  
While massless fields have been extensively studied, the inclusion of mass introduces a number of qualitatively new phenomena.  
\begin{itemize}
    \item Massive bosonic fields may give rise to characteristic signals in pulsar timing array (PTA) observations through the coupling of long-wavelength modes to the background spacetime \cite{Konoplya:2023fmh}.
\item Even fields that are intrinsically massless may acquire effective mass terms in certain contexts, such as brane-world scenarios or under strong electromagnetic fields \cite{Seahra:2004fg,Ishihara:2008re,Davlataliev:2024mjl,Konoplya:2007yy,Konoplya:2008hj,Wu:2015fwa,Kokkotas:2010zd}.  
\item The time-domain evolution of massive perturbations exhibits qualitatively different late-time behavior: instead of the power-law tails typical of massless fields, massive perturbations display oscillatory inverse-power tails \cite{Koyama:2001qw,Konoplya:2002wt,Jing:2004zb,Moderski:2001tk,Rogatko:2007zz}.  
\item In certain parameter regimes, long-lived and even arbitrarily long-lived quasinormal modes (the so-called quasiresonances) can occur \cite{Ohashi:2004wr, Konoplya:2017tvu,Bolokhov:2024bke,Zinhailo:2024jzt,Lutfuoglu:2025hwh,Bolokhov:2023bwm,Bolokhov:2023ruj,Skvortsova:2025cah,Skvortsova:2024eqi,Lutfuoglu:2025bsf,Churilova:2019qph,Dubinsky:2025bvf}, though their existence is not universal and should be verified for each specific background.  
\end{itemize}
Thus, the analysis of massive fields reveals rich dynamical phenomena and offers an additional observational channel for probing both the properties of black holes and the nature of quantum corrections near the horizon.

While several works have analyzed quasinormal oscillations of \textit{massless} fields in the Hayward background~\cite{Konoplya:2022hll,Malik:2024tuf,Konoplya:2023ppx,DuttaRoy:2022ytr,Bolokhov:2025egl}, the quasinormal behavior of \textit{massive} perturbations has so far remained unstudied, except the work~\cite{Lin:2013ofa}, where the data was represented only in the near-eikonal regime and small masses of the field, thereby not reflecting essential features of the spectrum of massive fields. The grey-body factors have not been studied for this case to the best of our knowledge. 

\vspace{1em}
In this work, we study quasinormal modes and grey-body factors of a massive scalar field in the background of the Hayward black hole considered as an effective metric emerging from Asymptotically Safe gravity.  
Our goal is to determine how the quantum-corrected geometry modifies the quasinormal spectrum and transmission coefficients, and to establish whether long-lived modes arise within physically relevant ranges of parameters.  
The results presented here may contribute to understanding the observational signatures of regular and quantum-corrected black holes.

The paper is organized as follows.  
In Sec. II, we briefly review the Hayward spacetime, emphasizing its interpretation both as a regular black hole and as an effective geometry emerging from Asymptotically Safe gravity.  
Section III formulates the wave equation for a massive scalar field as well as the numerical techniques employed for calculating the quasinormal frequencies—namely, the WKB method with Padé approximants and the time-domain integration followed by Prony analysis.  
In Sec. IV, we present and discuss the obtained quasinormal spectra, focusing on the influence of the field mass on the oscillation frequency and damping rate, and we identify the emergence of long-lived modes.  
Section V analyzes grey-body factors, including their behavior with respect to the field mass and multipole number, as well as to the verification of the correspondence between grey-body factors and quasinormal modes.  
Finally, the main results and their physical implications are summarized in Sec.V.

\section{The Hayward Spacetime}\label{sec:metric}

One of the long-standing problems in black-hole physics is the resolution of curvature singularities that inevitably arise in classical general relativity.  
Among the earliest and most influential proposals addressing this issue is the Hayward geometry~\cite{Hayward:2005gi}, which provides a simple yet self-consistent example of a nonsingular black hole with a regular core.  
This metric is part of a broader class of \emph{regular black holes}, characterized by a smooth interpolation between the standard Schwarzschild asymptotics at large radii and a de~Sitter–type interior near the origin.  
In such models, the singularity is replaced by a region of finite curvature, while an event horizon continues to exist, thereby preserving the essential causal structure of a black hole.

\vspace{1em}
The Hayward spacetime is described by the static, spherically symmetric line element
\begin{equation}
ds^2 = -f(r)\,dt^2 + \frac{dr^2}{f(r)} + r^2\!\left(d\theta^2 + \sin^2\!\theta\, d\phi^2\right),
\end{equation}
where the lapse function is given by
\begin{equation}\label{fr}
f(r) = 1 - \frac{2 M r^2}{r^3 + 2 M l^2}.
\end{equation}
Here $M$ is the black hole mass and $l$ determines the characteristic length scale at which non-classical or quantum-gravity effects become relevant.  
For radii much larger than $(M l^2)^{1/3}$, the function $f(r)$ approaches the Schwarzschild form, $f(r) \simeq 1 - 2M/r$,  
while near the center one finds $f(r) \simeq 1 - r^2/l^2$, corresponding to a de~Sitter core with finite curvature.  
Hence, the Hayward metric provides a continuous transition between the classical and quantum-dominated regimes.

An intriguing feature of the Hayward geometry is the close connection between its ultraviolet regularization scale and its horizon thermodynamics.  
The same parameter $l$ that determines the de~Sitter curvature at the core also fixes the near-extremal surface gravity, so that the onset of extremality corresponds to a balance between the inner (quantum) and outer (classical) scales of the metric~\cite{Hayward:2005gi,DeLorenzo:2014pta}.  
In this sense, the Hayward black hole exhibits a duality between the regular core and horizon thermodynamics, unlike other regular models such as those of Bardeen or Dymnikova.  
Moreover, if one requires that the stress–energy tensor satisfy the weak energy condition everywhere, the Hayward solution reaches the maximal allowed ratio $l/M \approx 1.06$ beyond which no event horizon forms~\cite{Balart:2014cga,Neves:2014aba}.  
This bound defines the transition between regular black holes and horizonless compact objects, highlighting the delicate balance between curvature regularization and energy conditions that distinguishes the Hayward spacetime within the family of regular black-hole metrics.

\vspace{1em}
Initially, the parameter $l$ was introduced on phenomenological grounds to mimic possible quantum backreaction or exotic matter that regularizes the singularity.  
However, subsequent theoretical developments revealed that this metric may naturally arise from more fundamental considerations.  
In particular, Held, Gold, and Eichhorn~\cite{Held:2019xde} demonstrated that, after a suitable redefinition of parameters, a function of the same form appears as an effective solution within the framework of Asymptotically Safe (AS) gravity.  
In this setting the metric function can be written as
\begin{equation}
f(r) = 1 - \frac{2 r^2 / M^2}{r^3 / M^3 + \gamma},
\end{equation}
where $\gamma$ quantifies the strength of the quantum-gravity corrections.  
The condition
\[
\gamma \lesssim \frac{32}{27}
\]
ensures that an event horizon exists and prevents the appearance of a naked singularity.  
In the AS scenario, the geometry emerges by relating the renormalization-group scale $k$ to curvature invariants, such as the Kretschmann scalar, leading to a scale-dependent effective metric.  
Different prescriptions for the cutoff identification produce a family of quantum-corrected black-hole solutions, including those proposed by Bonanno and Reuter~\cite{Bonanno:2000ep} and by Platania~\cite{Platania:2019kyx}.  
Quasinormal modes of test fields have been examined in several of these AS-inspired geometries, whereas gravitational perturbations have so far been explored only in a few cases, such as the Bonanno–Reuter spacetime~\cite{Bolokhov:2025lnt} and models describing quantum-corrected collapse~\cite{Bonanno:2023rzk,Shi:2025gst}.

\vspace{1em}
Therefore, the Hayward spacetime is significant in two key respects. First, it provides a paradigmatic example of a regular black hole that resolves the central singularity by means of an effective de~Sitter core.  
Second, it can be interpreted as an emergent, quantum-corrected geometry within the Asymptotically Safe framework.  
This dual interpretation makes the Hayward metric an ideal setting for investigating how quantum modifications of gravity influence observable quantities such as quasinormal spectra, grey-body factors, and the stability of the resulting black-hole configurations.

\begin{figure}
\resizebox{\linewidth}{!}{\includegraphics{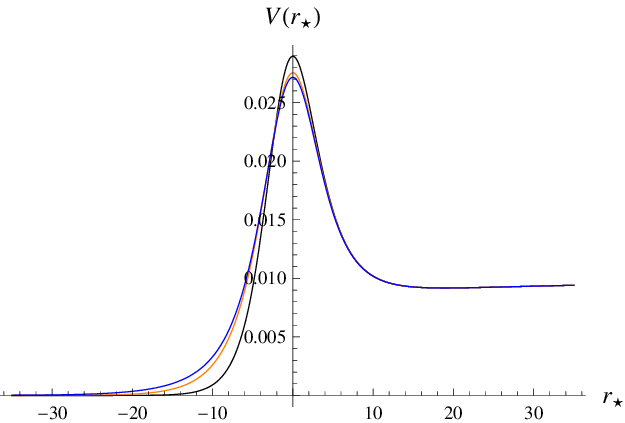}}
\caption{Potential as a function of the tortoise coordinate of the $\ell=0$, $\mu=0.1$ scalar field for the Hayward black hole ($M=1$): $\gamma=0$ (black)  (green) $\gamma=0.9$ (orange), and $\gamma=1.1$ (blue).}\label{fig:potentials1}
\end{figure}

\begin{figure}
\resizebox{\linewidth}{!}{\includegraphics{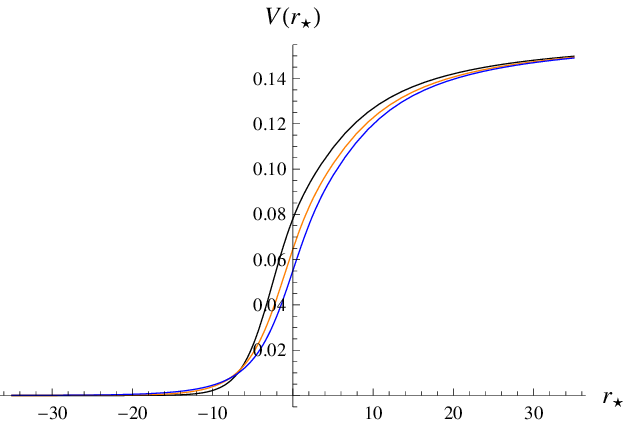}}
\caption{Potential as a function of the tortoise coordinate of the $\ell=0$, $\mu=0.4$ scalar field for the Hayward black hole ($M=1$): $\gamma=0$ (black)  $\gamma=0.9$ (orange), and $\gamma=1.1$ (blue). One can see that at sufficiently large mass of the field, the effective potential does not have a peak.}\label{fig:potentials2}
\end{figure}

\begin{table*}
\begin{tabular}{c c c c c}
\hline
$\mu$ & WKB6-Pade4 & WKB7-Pade4 & relative error $Re (\omega)$ & relative error $Im (\omega)$  \\
\hline
$0$ & $0.110839-0.088116 i$ & $0.110340-0.089232 i$ & $0.450\%$ & $1.27\%$\\
$0.01$ & $0.110877-0.088086 i$ & $0.110419-0.089179 i$ & $0.413\%$ & $1.24\%$\\
$0.05$ & $0.113598-0.089056 i$ & $0.112374-0.087494 i$ & $1.08\%$ & $1.75\%$\\
$0.1$ & $0.117615-0.079742 i$ & $0.116187-0.081219 i$ & $1.21\%$ & $1.85\%$\\
$0.15$ & $0.120916-0.070628 i$ & $0.121269-0.071191 i$ & $0.292\%$ & $0.796\%$\\
\hline
\end{tabular}
\caption{Quasinormal modes of the $\ell=0$, $n=0$ scalar field for the Hayward black hole calculated using the 6th and 7th orders WKB formula with $m=4$ Padé approximant for various values of $\mu$; $M=1$, $\gamma=1.18$. The expected relative error is much smaller than the effect.}
\end{table*}

\begin{table*}
\begin{tabular}{c c c c c}
\hline
$\mu$ & WKB6-Pade4:6 & WKB7-Pade4 & relative error $Re (\omega)$ & relative error $Im (\omega)$  \\
\hline
$0$ & $0.307461-0.082194 i$ & $0.307464-0.082201 i$ & $0.00096\%$ & $0.0086\%$\\
$0.01$ & $0.307503-0.082178 i$ & $0.307506-0.082185 i$ & $0.00093\%$ & $0.0084\%$\\
$0.1$ & $0.311568-0.080557 i$ & $0.311624-0.080554 i$ & $0.0178\%$ & $0.0039\%$\\
$0.2$ & $0.324150-0.075221 i$ & $0.324192-0.075242 i$ & $0.0129\%$ & $0.0282\%$\\
$0.3$ & $0.345274-0.064969 i$ & $0.345346-0.064999 i$ & $0.0208\%$ & $0.0471\%$\\
$0.4$ & $0.375140-0.047589 i$ & $0.374978-0.048212 i$ & $0.0431\%$ & $1.31\%$\\
\hline
\end{tabular}
\caption{Quasinormal modes of the $\ell=1$, $n=0$ scalar field for the Hayward black hole calculated using the 6th and 7th orders WKB formula with $m=4$ Padé approximant for various values of $\mu$; $M=1$, $\gamma=1.18$.}
\end{table*}

\begin{table*}
\begin{tabular}{c c c c c}
\hline
$\mu$ & WKB6-Pade4 & WKB7-Pade4 & relative error $Re (\omega)$ & relative error $Im (\omega)$  \\
\hline
$0$ & $0.509665-0.081743 i$ & $0.509667-0.081745 i$ & $0.00046\%$ & $0.0028\%$\\
$0.01$ & $0.509692-0.081737 i$ & $0.509694-0.081739 i$ & $0.00046\%$ & $0.0028\%$\\
$0.1$ & $0.512393-0.081119 i$ & $0.512393-0.081121 i$ & $0.00017\%$ & $0.0013\%$\\
$0.2$ & $0.520618-0.079189 i$ & $0.520628-0.079189 i$ & $0.00195\%$ & $0.0003\%$\\
$0.3$ & $0.534534-0.075767 i$ & $0.534542-0.075767 i$ & $0.00144\%$ & $0.0006\%$\\
$0.4$ & $0.554426-0.070514 i$ & $0.554433-0.070513 i$ & $0.00121\%$ & $0.0009\%$\\
$0.5$ & $0.580741-0.062842 i$ & $0.580746-0.062841 i$ & $0.00075\%$ & $0.0016\%$\\
$0.6$ & $0.614078-0.051712 i$ & $0.613984-0.051733 i$ & $0.0154\%$ & $0.0399\%$\\
$0.7$ & $0.654895-0.035822 i$ & $0.654557-0.035513 i$ & $0.0517\%$ & $0.862\%$\\
\hline
\end{tabular}
\caption{Quasinormal modes of the $\ell=2$, $n=0$ scalar field for the Hayward black hole calculated using the 6th and 7th orders WKB formula with $m=4$ Padé approximant for various values of $\mu$; $M=1$, $\gamma=1.18$.}
\end{table*}

\begin{table*}
\begin{tabular}{c c c c c}
\hline
$\mu$ & WKB6-Pade4 &WKB7-Pade4 & relative error $Re (\omega)$ & relative error $Im (\omega)$  \\
\hline
$0$ & $0.487149-0.248544 i$ & $0.487310-0.248333 i$ & $0.0330\%$ & $0.0849\%$\\
$0.01$ & $0.487169-0.248527 i$ & $0.487330-0.248317 i$ & $0.0330\%$ & $0.0849\%$\\
$0.1$ & $0.489183-0.246913 i$ & $0.489348-0.246706 i$ & $0.0338\%$ & $0.0838\%$\\
$0.2$ & $0.495235-0.241958 i$ & $0.495396-0.241733 i$ & $0.0325\%$ & $0.0932\%$\\
$0.3$ & $0.505133-0.233485 i$ & $0.505252-0.233214 i$ & $0.0236\%$ & $0.116\%$\\
$0.4$ & $0.518546-0.221189 i$ & $0.518664-0.220867 i$ & $0.0228\%$ & $0.146\%$\\
$0.5$ & $0.534939-0.204732 i$ & $0.535131-0.204501 i$ & $0.0360\%$ & $0.113\%$\\
$0.6$ & $0.548406-0.182857 i$ & $0.552711-0.183863 i$ & $0.785\%$ & $0.550\%$\\
\hline
\end{tabular}
\caption{Quasinormal modes of the $\ell=2$, $n=1$ scalar field for the Hayward black hole calculated using the 6th and 7th orders WKB formula with $m=4$ Padé approximant for various values of $\mu$; $M=1$, $\gamma=1.18$.}
\end{table*}

\begin{table*}
\begin{tabular}{c c c c c}
\hline
$\mu$ & WKB6Pade4:6 & WKB7-Pade4 & relative error $Re (\omega)$ & relative error $Im (\omega)$  \\
\hline
$0$ & $0.442963-0.425943 i$ & $0.442422-0.425480 i$ & $0.122\%$ & $0.109\%$\\
$0.01$ & $0.442973-0.425923 i$ & $0.442432-0.425460 i$ & $0.122\%$ & $0.109\%$\\
$0.1$ & $0.443950-0.423879 i$ & $0.443443-0.423484 i$ & $0.114\%$ & $0.0933\%$\\
$0.2$ & $0.446651-0.417333 i$ & $0.446924-0.417251 i$ & $0.0610\%$ & $0.0197\%$\\
$0.3$ & $0.450800-0.406520 i$ & $0.450854-0.406709 i$ & $0.0120\%$ & $0.0464\%$\\
$0.4$ & $0.458087-0.392622 i$ & $0.458006-0.390807 i$ & $0.0175\%$ & $0.462\%$\\
$0.5$ & $0.472093-0.374536 i$ & $0.471121-0.375452 i$ & $0.206\%$ & $0.244\%$\\
\hline
\end{tabular}
\caption{Quasinormal modes of the $\ell=2$, $n=2$ scalar field for the Hayward black hole calculated using the 6th and 7th orders WKB formula with $m=4$ Padé approximant for various values of $\mu$; $M=1$, $\gamma=1.18$.}
\end{table*}

\begin{table*}
\begin{tabular}{c c c c c}
\hline
$\gamma$ & WKB6Pade4:6 & WKB7-Pade4 & relative error $Re (\omega)$ & relative error $Im (\omega)$  \\
\hline
$0$ & $0.486803-0.095675 i$ & $0.486804-0.095675 i$ & $0.00018\%$ & $0.0005\%$\\
$0.3$ & $0.492511-0.093258 i$ & $0.492513-0.093255 i$ & $0.00026\%$ & $0.0030\%$\\
$0.5$ & $0.496617-0.091313 i$ & $0.496615-0.091308 i$ & $0.00044\%$ & $0.0064\%$\\
$0.7$ & $0.500980-0.089000 i$ & $0.500977-0.088997 i$ & $0.00056\%$ & $0.0035\%$\\
$0.9$ & $0.505601-0.086198 i$ & $0.505600-0.086197 i$ & $0.00029\%$ & $0.0014\%$\\
$1.1$ & $0.510432-0.082734 i$ & $0.510433-0.082735 i$ & $0.00023\%$ & $0.0011\%$\\
$1.18$ & $0.512393-0.081119 i$ & $0.512393-0.081121 i$ & $0.00017\%$ & $0.0013\%$\\
\hline
\end{tabular}
\caption{Quasinormal modes of the $\ell=2$, $n=2$ scalar field for the Hayward black hole calculated using the 6th and 7th orders WKB formula with $m=4$ Padé approximant for various values of $\gamma$; $M=1$, $\mu=0.1$. The expected relative error is much smaller than the effect.}
\end{table*}

\section{Numerical Methods for Calculating Quasinormal Modes}

\subsection{Wave equation and boundary conditions}

The dynamics of a massive scalar field $\Phi$ of mass $\mu$ in a static, spherically symmetric spacetime are governed by the Klein--Gordon equation
\begin{equation}
\frac{1}{\sqrt{-g}}\partial_\mu\!\left(\sqrt{-g}\, g^{\mu\nu}\partial_\nu \Phi\right) - \mu^2 \Phi = 0.
\end{equation}
After separation of variables
\begin{equation}
\Phi(t,r,\theta,\phi) = \frac{\psi_\ell(r)}{r}\, Y_{\ell m}(\theta,\phi)\, e^{-i \omega t},
\end{equation}
the radial function $\psi_\ell(r)$ satisfies a Schrödinger-like equation of the form
\begin{equation}
\frac{d^2 \psi_\ell}{d r_*^2} + \left[\omega^2 - V_\ell(r)\right] \psi_\ell = 0,
\label{wave_eq}
\end{equation}
where the tortoise coordinate $r_*$ is defined by  $dr_*/dr = 1/f(r)$, and the effective potential reads
\begin{equation}
V_\ell(r) = f(r)\left[\frac{\ell(\ell+1)}{r^2} + \frac{f'(r)}{r} + \mu^2\right].
\label{potential}
\end{equation}

For an asymptotically flat black hole, the potential $V_\ell(r)$ has a single maximum outside the event horizon and decays exponentially at both boundaries:
\[
V_\ell(r\to r_h)\to 0, \qquad V_\ell(r\to\infty)\to \mu^2.
\]
Quasinormal modes are defined by purely ingoing and outgoing boundary conditions,
\begin{equation}
\psi_\ell(r) \sim
\begin{cases}
e^{-i \omega r_*}, & r_* \to -\infty \quad (r \to r_h), \\[4pt]
e^{+i k r_*}, & r_* \to +\infty \quad (r \to \infty),
\end{cases}
\label{boundary}
\end{equation}
where $k = \sqrt{\omega^2 - \mu^2}$ and $\re{k}$ has the same sign as $\re{\omega}$.

\subsection{WKB and Padé–improved approximation}

For potentials of the form~(\ref{potential}) that exhibit a single peak, the Wentzel–Kramers–Brillouin (WKB) method provides accurate approximations for low overtones.  
Expanding $V(r)$ near its maximum $r_0$ and imposing the boundary conditions~(\ref{boundary}), one obtains the standard WKB quantization condition~\cite{Schutz:1985km,Iyer:1986np,Konoplya:2003ii}
\begin{equation}
\frac{i\,(\omega^2 - V_0)}{\sqrt{-2 V_0''}} - \sum_{j=2}^{N} \Lambda_j = n + \frac{1}{2},
\label{WKB_condition}
\end{equation}
where $V_0 = V(r_0)$, $V_0'' = d^2 V / dr_*^2 |_{r_0}$, $n=0,1,2,\dots$ is the overtone number, and $\Lambda_j$ denote higher-order WKB corrections up to the chosen order $N$.

To enhance convergence and accuracy, it is convenient to recast~(\ref{WKB_condition}) into a rational Padé form~\cite{Konoplya:2019hlu,Matyjasek:2017psv},
$$\sout{\omega^2 = V_0 - i \sqrt{-2 V_0''}\, \mathcal{P}_{m,n}\!\left(n+\frac{1}{2}\right),}$$
\begin{equation}
\omega^2 = \mathcal{P}_{m,n},
\label{WKB_Pade}
\end{equation}
where $\mathcal{P}_{m,n}$ denotes the $(m,n)$ Padé approximant constructed from the WKB expansion series.  
The Padé–WKB method yields highly accurate results for the fundamental and first few overtones, typically with relative errors below $10^{-3}$ for $\ell \ge 2$ in asymptotically flat or de Sitter spacetimes~\cite{Qian:2022kaq, Dubinsky:2024mwd, Malik:2024elk, Cuyubamba:2016cug, Konoplya:2024lch, Churilova:2021tgn, Skvortsova:2023zca, Stuchlik:2025mjj, Konoplya:2020hyk, Malik:2024qsz, Konoplya:2013sba,  Momennia:2022tug, Dubinsky:2024gwo, Malik:2024nhy, Dubinsky:2024aeu, Dubinsky:2025fwv, Aneesh:2018hlp, Zhao:2023itk, Zhao:2023tyo}. For the massive field, the potential~(\ref{potential}) is slightly modified, but the same formalism applies as long as the potential remains of the single-barrier type.

\subsection{Time-domain integration and Prony analysis}

To verify and complement the frequency-domain results, we also employ the time-domain integration method~\cite{Gundlach:1993tp} and extract quasinormal frequencies using the Prony technique~\cite{Konoplya:2011qq}.  
By rewriting Eq.~(\ref{wave_eq}) in null coordinates $u = t - r_*$ and $v = t + r_*$,
\begin{equation}
4 \frac{\partial^2 \psi}{\partial u\,\partial v} + V(r)\, \psi = 0,
\label{null_wave}
\end{equation}
the equation can be integrated numerically on a characteristic grid with the standard discretization
\begin{equation}
\psi(N) = \psi(W) + \psi(E) - \psi(S) - \frac{\Delta^2}{8}\, V(S)\,[\psi(W) + \psi(E)],
\end{equation}
where $N$, $W$, $E$, and $S$ denote the grid points $(u+\Delta,v+\Delta)$, $(u+\Delta,v)$, $(u,v+\Delta)$, and $(u,v)$, respectively.  
A Gaussian pulse is typically used as the initial condition on the $u$ and $v$ axes.  
The resulting waveform $\psi(t,r_{\mathrm{obs}})$ at a fixed observer location exhibits a ringdown phase that can be fitted as a superposition of damped sinusoids,
\begin{equation}
\psi(t) = \sum_{j=1}^{p} C_j\, e^{-i \omega_j t},
\label{prony_fit}
\end{equation}
where $p$ is the number of dominant modes.  
The complex frequencies $\omega_j$ are extracted by solving the corresponding Prony system, which provides excellent precision for the fundamental mode when applied to the later portion of the ringdown signal~\cite{Skvortsova:2024wly, Kodama:2009bf, Lutfuoglu:2025ljm, Skvortsova:2023zmj, Zhao:2022gxl,  Skvortsova:2024atk, Bolokhov:2022rqv, Konoplya:2014lha, del-Corral:2022kbk, Lutfuoglu:2025pzi,  Lutfuoglu:2025qkt, Bolokhov:2024ixe, Lutfuoglu:2025ohb, Wang:2024gzr}.

\subsection{Accuracy and applicability}

The Padé-improved WKB approximation and the time-domain–Prony analysis complement each other.  
The former gives a fast and accurate estimation of $\omega$ for intermediate and high multipoles ($\ell \ge 2$), while the latter remains reliable even when the potential is distorted by the mass term or quantum corrections, as in the Hayward geometry.  
Consistency between the two approaches serves as an important verification of the obtained quasinormal spectra. We will measure all the dimensional quantities in units of the black-hole mass, so from now on we choose $M=1$.

\begin{figure}
\resizebox{\linewidth}{!}{\includegraphics{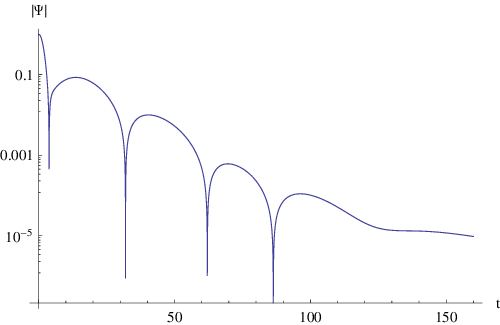}}
\caption{Semi-lograrithmic time-domain profile for $\gamma=1.18$, $\ell=0$, $\mu=0$, $M=1$. The Prony method allows one to extract the fundamental mode $\omega = 0.110201 - 0.0881366 i$, and the 6th order WKB method gives $\omega = 0.110839 - 0.088116  i$. The difference between the quasinormal modes obtained by the two methods is $0.000638 + 0.000021 i$, which is considerably smaller than one tenth of a percent.}\label{fig:TDL0mu0}
\end{figure}

\begin{figure}
\resizebox{\linewidth}{!}{\includegraphics{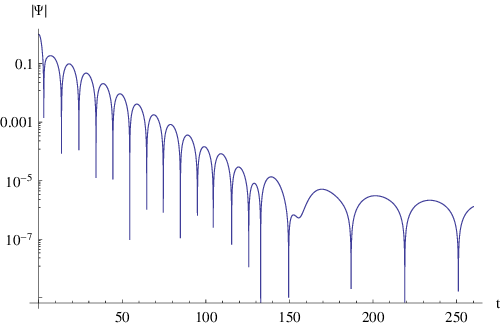}}
\caption{Semi-lograrithmic time-domain profile for $\gamma=1.18$, $\ell=1$, $\mu=0.1$, $M=1$. The Prony method allows one to extract the fundamental mode $\omega = 0.311752 - 0.079697 i$, and the 6th order WKB method gives $\omega = 0.311568 - 0.080557  i$. The difference between the quasinormal modes obtained by the two methods is $-0.000184 - 0.0008598 i$, which is considerably smaller than one tenth of a percent. After the ringdown phase one can see the beginning of the intermediate oscillatory tail with the power-law envelope.}\label{fig:TDL1mu01}
\end{figure}

\begin{figure}
\resizebox{\linewidth}{!}{\includegraphics{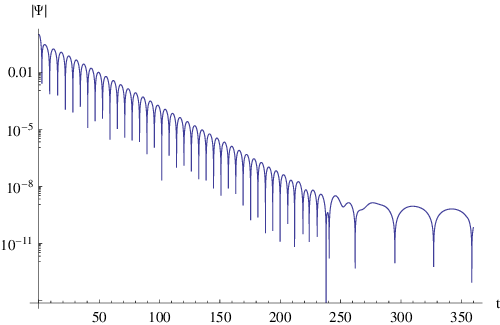}}
\caption{Semi-lograrithmic time-domain profile for $\gamma=1.18$, $\ell=2$, $\mu=0.1$, $M=1$. The Prony method allows one to extract the fundamental mode $\omega = 0.512372 - 0.0811234 i$, and the 6th order WKB method gives $\omega = 0.512393 - 0.081119 i$. The difference between the quasinormal modes obtained by the two methods $0.0000212532 + 4.43302*10^-6 I$ is negligible. After the ringdown phase one can see the beginning of the intermediate oscillatory tail with the power-law envelope.}\label{fig:TDL2mu01}
\end{figure}

\section{Quasinormal modes}

In Fig.~\ref{fig:TDL0mu0} (massless case, $\ell=0$), the semi-logarithmic time-domain profile yields a fundamental mode from the Prony fit, $\omega = 0.110201 - 0.0881366 i$, which agrees closely with the 6th-order WKB value, $\omega = 0.110839 - 0.088116 i$.  
The difference, $0.000638 + 0.000021\,i$, is well below one-tenth of a percent in both real and imaginary parts, indicating very good consistency between the two independent methods in the $\mu=0$ baseline.

For $\mu=0.1$, the profiles in Fig.~\ref{fig:TDL1mu01} ($\ell=1$) and Fig.~\ref{fig:TDL2mu01} ($\ell=2$) show similarly close agreement.  
In Fig.~\ref{fig:TDL1mu01}, the Prony result $\omega = 0.311752 - 0.079697 i$ is near the WKB value $\omega = 0.311568 - 0.080557 i$, with a small difference of $-0.000184 - 0.0008598\,i$.  
In Fig.~\ref{fig:TDL2mu01}, the match is even tighter: $\omega_{\text{Prony}} = 0.512372 - 0.0811234 i$ versus $\omega_{\text{WKB}} = 0.512393 - 0.081119 i$, and the residual $2.12532\times 10^{-5} + 4.43302\times 10^{-6} i$ is negligible.  
In both massive cases (Figs.~\ref{fig:TDL1mu01} and~\ref{fig:TDL2mu01}), after the ringdown stage the waveform exhibits the onset of an intermediate \textit{oscillatory tail with a power-law envelope}, as noted in the captions, consistent with the characteristic late-time behavior of massive fields in asymptotically flat backgrounds.

The data presented in Tables~I--VI allow us to trace in detail how both the field mass and the Hayward parameter affect the quasinormal spectrum.  
For the fundamental modes ($n=0$) of the lowest multipoles, the real oscillation frequency $\re{\omega}$ grows monotonically with the field mass $\mu M$, while the damping rate $|\im{\omega}|$ decreases.  
This trend is clearly seen for $\ell=0,1,2$ in Tables~I--III, where $\re{\omega}$ increases from approximately $0.11$ to $0.65$ as $\mu M$ rises, whereas $|\im{\omega}|$ is reduced by roughly a factor of two to three over the same range.  
The difference between the Padé-improved sixth- and seventh-order WKB approximations remains at the level of $10^{-3}$--$10^{-2}$ in relative units, confirming that the numerical uncertainty is much smaller than the physical variation of the spectrum with $\mu$.  
For large field masses, the imaginary part tends to zero, marking the transition toward the regime of long-lived or quasi-resonant states, although no instability is observed within the considered parameter domain.  

A similar monotonic behavior is observed for the overtones.  
As shown in Tables~IV and~V, for the first and second overtones ($n=1,2$) of the $\ell=2$ mode, the oscillation frequency increases and the damping rate decreases with growing $\mu M$, following the same pattern as the fundamental mode but with larger absolute values of $|\mathrm{Im}(\omega)|$, as expected.  
The agreement between the values obtained at the 6th and 7th WKB orders remains good for all overtones, although the relative difference becomes slightly larger at high $\mu M$, where the potential barrier becomes shallower and the WKB expansion less accurate.  Beyond the highest masses shown in the tables, the effective potential loses its single-peak shape and the standard WKB formalism can no longer be applied.  Therefore, the quasi-resonances regime cannot be achieved via the WKB approach in principle, although the extrapolation to larger $\mu$ show that the quasi-resonances should exist in the spectrum.

The influence of the Hayward (or Asymptotic Safety) parameter $\gamma$ is summarized in Table~VI.  
An increase in $\gamma$ leads to a mild upward shift of $\mathrm{Re}(\omega)$ and a moderate suppression of the damping rate, $|\mathrm{Im}(\omega)|$.  
This indicates that the presence of a regular core or quantum-corrected geometry tends to make the oscillations slightly more stable, but the overall effect of $\gamma$ is much weaker than that of the field mass. This mild dependence on the quantum parameter is clear, because the effective potential deviates from that of the Schwarzschild spacetime, mainly only near the event horizon. This produce outburst of overtones, but guarantees weak sensitivity of the fundamental mode \cite{Konoplya:2023aph}.

\section{Grey-body factors for massive fields}

\begin{figure*}
\resizebox{\linewidth}{!}{\includegraphics{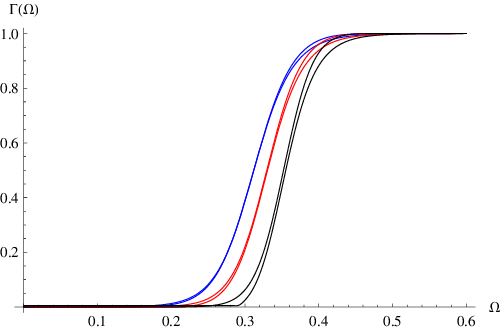}\includegraphics{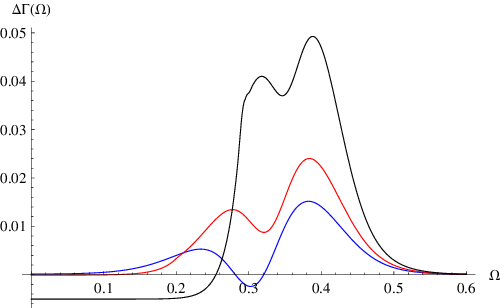}}
\caption{Grey-body factors for $\ell=1$ perturbations at $\gamma=1.18$, $M=1$, and various $\mu$: $\mu=0$ (blue), $\mu=0.2$ (red) and $\mu=0.3$ (black). We can see that when $\mu$ is increased, the emission of lower frequencies is suppressed.}\label{fig:GBFL1}
\end{figure*}

\begin{figure*}
\resizebox{\linewidth}{!}{\includegraphics{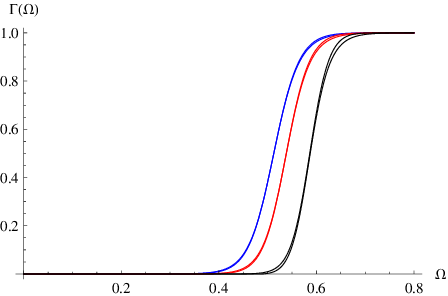}\includegraphics{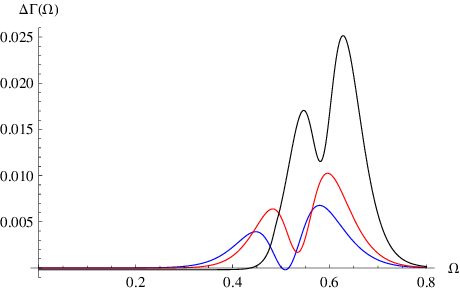}}
\caption{Grey-body factors for $\ell=2$ perturbations at $\gamma=1.18$, $M=1$, and various $\mu$: $\mu=0$ (blue), $\mu=0.3$ (red) and $\mu=0.5$ (black). We can see that when $\mu$ is increased, the emission of lower frequencies is suppressed. The relative error of the correspondence decreases strongly as $\ell$ grows, as can be seen via comparison with fig. \ref{fig:GBFL1}.}\label{fig:GBFL2}
\end{figure*}

In an asymptotically flat spacetime, the propagation of a massive scalar field obeys the radial equation
\begin{equation}
\frac{d^2 \psi_\ell}{dr_*^2} + \left[\omega^2 - V_\ell(r)\right] \psi_\ell = 0,
\end{equation}
with the effective potential $V_\ell(r)$ defined by Eq.~(\ref{potential}).  
At the event horizon ($r_* \to -\infty$) and at spatial infinity ($r_* \to +\infty$), the potential vanishes asymptotically, and the physically acceptable solutions satisfy
\begin{equation}
\psi_\ell \sim
\begin{cases}
T_\ell(\omega)\, e^{-i\omega r_*}, & r_* \to -\infty, \\[4pt]
e^{-i k r_*} + R_\ell(\omega)\, e^{+i k r_*}, & r_* \to +\infty,
\end{cases}
\label{gbf_boundaries}
\end{equation}
where $k=\sqrt{\omega^2-\mu^2}$, and the coefficients $R_\ell$ and $T_\ell$ correspond to reflection and transmission amplitudes, respectively.  
The \emph{grey-body factor} (GBF) measures the probability that an outgoing mode emitted near the horizon penetrates the potential barrier and reaches infinity,
\begin{equation}
\Gamma_\ell(\omega) = |T_\ell(\omega)|^2 = 1 - |R_\ell(\omega)|^2,
\end{equation}
and determines the modification of the Hawking radiation spectrum relative to that of a perfect black body.  
For $\mu \neq 0$, the field acquires an effective mass barrier at large $r$, so that $\Gamma_\ell(\omega)$ vanishes when $\omega < \mu$ and approaches unity in the high-frequency limit.  
The mass term therefore introduces a frequency cutoff and suppresses low-energy transmission, leading to a shift of the emission peak toward higher frequencies.

\subsection{Computation via the WKB approximation and the QNM correspondence}

The grey-body factors may be computed semi-analytically via the WKB approach developed for potential barriers with a single maximum~\cite{Iyer:1986np,Konoplya:2003ii,Konoplya:2019hlu}.  
Expanding the potential around its peak $r_0$ and matching the asymptotic solutions across the turning points yields
\begin{equation}
\Gamma_\ell(\omega)
 = \left(1 + \exp\!\left[2\pi\,\frac{(\omega^2 - V_0)}{\sqrt{-2 V_0''}}\right]\right)^{-1},
\label{gbf_wkb}
\end{equation}
where $V_0 = V_\ell(r_0)$ and $V_0'' = d^2V/dr_*^2|_{r_0}$.  
Higher-order WKB corrections and Padé resummation can be incorporated to improve accuracy in the near-barrier region, providing reliable results for $\ell \ge 1$ and zero or moderate masses $\mu M \lesssim 1$ (see recent examples in \cite{Konoplya:2023ahd, Dubinsky:2025ypj, Dubinsky:2025nxv, Dubinsky:2024vbn, Lutfuoglu:2025eik, Lutfuoglu:2025blw, Konoplya:2021ube, Lutfuoglu:2025ldc, Lutfuoglu:2025kqp} for recent examples).

An alternative and particularly efficient technique exploits the correspondence between quasinormal modes and grey-body factors in the eikonal regime~\cite{Konoplya:2024lir,Konoplya:2024vuj}.  
Since both quantities originate from the same scattering potential, the reflection and transmission coefficients can be approximated in terms of the complex quasinormal frequency $\omega_0 = \omega_R - i\,\omega_I$ of the fundamental mode.  
In this approach, the transmission probability assumes a resonant form
\begin{equation}
\Gamma_\ell(\omega) \simeq 
\frac{1}{1 + \exp\!\left[2\pi\,\frac{\omega^2 - \omega_R^2}{\omega_I\,\omega_R}\right]},
\label{gbf_qnm}
\end{equation}
which shows that the grey-body spectrum peaks near $\omega \simeq \omega_R$, with the resonance width determined by the damping rate $\omega_I$.  
Hence, the maxima of $\Gamma_\ell(\omega)$ correspond to the oscillation frequencies of the dominant quasinormal mode, while their widths reflect the imaginary parts of the same modes.  

According to \cite{Konoplya:2024lir,Konoplya:2024vuj}, the origin of this expression can be seen directly from the leading-order WKB correspondence between the potential peak and the fundamental quasinormal frequency. We start from the standard WKB expression for the transmission probability through a potential barrier with a single maximum \ref{gbf_wkb}. For the fundamental mode ($n=0$), the WKB quantization condition gives
\[
\omega_0^2 \simeq V_0-\frac{i}{2}\sqrt{-2V_0''}.
\]
Writing $\omega_0=\omega_R-i\omega_I$, one finds
\[
\omega_0^2=(\omega_R-i\omega_I)^2
=\underbrace{\omega_R^2-\omega_I^2}_{\simeq\,\omega_R^2}
-2i\,\omega_R\omega_I .
\]
Matching real and imaginary parts yields
\[
V_0 \simeq \omega_R^2, \quad
\frac{1}{2}\sqrt{-2V_0''}=2\,\omega_R\omega_I
\;\Rightarrow\;
\sqrt{-2V_0''}=4\,\omega_R\omega_I .
\]
Substituting these relations into the transmission formula gives (\ref{gbf_qnm}).

This correspondence highlights the close connection between the black hole’s resonant response and its transmission properties: quasinormal modes describe the characteristic oscillations of the system, whereas grey-body factors quantify the probability for radiation to escape the potential barrier.  The agreement between $\Gamma_\ell(\omega)$ obtained from direct WKB computations and from Eq.~(\ref{gbf_qnm}) provides an important cross-check of numerical accuracy for both massive and massless perturbations. For this purpose we used the correspondence developed up to the third order beyond the above described eikonal limit \cite{Konoplya:2024lir,Konoplya:2024vuj}. The correspondence has already been tested in various contexts \cite{Bolokhov:2024otn, Malik:2025erb, Malik:2025dxn, Skvortsova:2024msa, Malik:2025qnr, Malik:2024cgb}, showing that it is precise in the eikonal regime  and reasonably accurate for lower multipole numbers $\ell$. At the same time the correspondence is based on the validity of the WKB approach  in the high $\ell$-regime and it can break down when the WKB approach breaks down in the eikonal regime \cite{Konoplya:2022gjp,Konoplya:2017wot,Bolokhov:2023dxq,Khanna:2016yow,Konoplya:2017lhs,Konoplya:2017ymp}.

\section{Conclusions}

While quasinormal modes of massless test fields in the Hayward spacetime have been extensively studied in the literature~\cite{Toshmatov:2019gxg,Konoplya:2022hll,DuttaRoy:2022ytr}, the spectrum of massive fields was considered only in Ref.~\cite{Lin:2013ofa}.  
However, that analysis presented results exclusively for the $\ell = 5$ multipole and for relatively small field masses, which are uncharacteristic for such large $\ell$.  
Consequently, the main feature of the massive-field spectrum—the presence of long-lived modes that continuously evolve into arbitrarily long-lived states (quasi-resonances)—was overlooked in that work.  
In this study, we presented a comprehensive analysis of the quasinormal spectrum and grey-body factors for massive scalar fields in the Hayward black hole background. We found that increasing the field mass markedly reduces the damping rate, pointing to the emergence of quasi-resonant modes at certain threshold masses.  
In the time domain, the usual exponentially decaying ringdown is replaced by oscillatory tails with a power-law envelope.   We further demonstrate that the correspondence between quasinormal modes and grey-body factors remains valid for massive fields, showing excellent accuracy for large multipole numbers, but gradually losing precision as the field mass increases or the multipole number decreases.

\begin{acknowledgments}
The author thanks R. A. Konoplya for useful discussions. 
The author acknowledges the University of Seville
for their support through the Plan-US of aid to Ukraine
\end{acknowledgments}

\bibliography{bibliography}

\end{document}